\documentclass{article}

\usepackage{PRIMEarxiv}
\usepackage{gensymb}
\usepackage[utf8]{inputenc} % allow utf-8 input
\usepackage[T1]{fontenc}    % use 8-bit T1 fonts
\usepackage{hyperref}       % hyperlinks
\usepackage{url}            % simple URL typesetting
\usepackage{booktabs}       % professional-quality tables
\usepackage{amsfonts}       % blackboard math symbols
\usepackage{nicefrac}       % compact symbols for 1/2, etc.
\usepackage{microtype}      % microtypography
\usepackage{lipsum}
\usepackage{fancyhdr}       % header
\usepackage{graphicx}       % graphics
\graphicspath{{media/}}     % organize your images and other figures under media/ folder

\title{Morphology Transition with Temperature and their Effect on Optical Properties of Colloidal MoS$_2$ Nanostructures}

\author{
  Simran Lambora, Asha Bhardwaj \\
  Department of Instrumentation and Applied Physics \\
  Indian Institute of Science Bangalore \\
  India\\
  \texttt{simranl@iisc.ac.in} \\
}

\begin{document}
\maketitle

\begin{abstract}
Morphology plays a crucial role in deciding the chemical and optical properties of nanomaterials due to confinement effects. We report the morphology transition of colloidal molybdenum disulfide (MoS$_2$) nanostructures, synthesized by one pot heat-up method, from mix of quantum dots (QDs) and nanosheets to predominantly nanorods by varying the synthesis reaction temperature from 90 to 160 \degree C. The stoichiometry and composition of the synthesized QDs, nanosheets and nanorods have been quantified to be MoS$_2$ using energy dispersive X-ray spectroscopy and X-ray photoelectron spectroscopy analysis. Nanostructure morphology transition due to variation in reaction temperature has resulted in photoluminescence quantum yield enhancement from zero to 4.4\% on increase in temperature from 90 to 120 \degree C. On further increase in temperature to 160 \degree C, a decrease in quantum yield to 2.63\% is observed. A red shift of $\approx$18 nm and $\approx$140 nm in the emission maxima and absorption edge respectively is observed for the synthesized nanostructures with increase in reaction temperature from 90 to 160 \degree C. The change in the quantum yield is attributed to the change in shape and hence confinement of charge carriers. To the best of our knowledge, first-time microscopic analysis of colloidal MoS$_2$ nanostructures shape and optical property variation with temperature explained by non-classical growth mechanism is presented.
\end{abstract}

% keywords can be removed

\section{Introduction}
During the last decade, nanostructured MoS$_2$ has attracted tremendous attention due to its intrinsic characteristics such as tunable electronic and optical properties, mechanical stability, and photochemical reactivity with applications such as field effect transistors, sensing, energy storage and catalysis\cite{samadi2018group,wang2012electronics,bazaka2019mos2,musselman2018research}. MoS$_2$ has layered S-Mo-S structure where Mo atom is covalently bonded to both S atoms with van Der Waals interaction between interlayers. The physical, chemical, electrical, and optical properties of MoS$_2$ depend on morphology and size of nanostructures due to quantum confinement effects. MoS$_2$ in bulk has indirect bandgap ($\approx$1 eV) which transforms to direct bandgap material in 2D ($\approx$1.85 eV)\cite{splendiani2010emerging,sundaram2013electroluminescence} and 0D ($\approx$4.95 eV)\cite{lambora2022uv,lambora2022highly}, along with enhancement in electronic, optical, catalytic and sensing characteristics due to confinement effects, opening a path for optoelectronic and biomedical applications such as optical sensing and bioimaging\cite{ayari2007realization,kwon2019monolayer,liu2016facile,lin2015colloidal,gao2013ferromagnetism,veeramalai2019highly,mishra2018highly,dong2016fluorescent,mukherjee2016novel}.

Methods such as liquid exfoliation and hydrothermal synthesis, have been used to fabricate MoS$_2$ QDs. However, these synthesis techniques suffer from shortcomings, such as long synthesis durations, low productivity and small QYs\cite{mukherjee2016novel,xu2015one,xu2019fabrication,gan2015quantum}, which are avoided in colloidal synthesis\cite{lambora2022uv,lambora2022highly,lin2015colloidal,lambora2023role,lambora2022optimization}. Further, controllability of size and morphology makes colloidal method a preferred route for nanocrystal formation where synthesis takes place in three sequential steps of monomer formation, nucleation, and growth of nuclei\cite{wen2021interplay,baek2021recent,xue2014crystal,thanh2014mechanisms,polte2015fundamental}. These monomers can nucleate either through classical condensation, leading to production of uniform sized nanoparticles through layer-by-layer crystal growth or through non-classical crystallization favouring particle attachment by aggregation, or oriented coagulation. Non-classical growth avoids the bottleneck of classical mechanism by circumventing constraint of crystal unit cell formation, enabling morphology transition by growth along specific (low energy) facet, coalescence or oriented attachment (OA) of nanoclusters which is not achievable by classical mechanism\cite{jehannin2019new}. Control over size and morphology in colloidal method can be done by varying reaction temperature, reaction time, capping agents, and precursor concentration\cite{madras2002transition,madras2003temperature,madras2004temperature}.

In this article, we report the first-time study of morphological transition of MoS$_2$ nanostructures and their optical properties relation with reaction temperature (T$_R$). Different morphology MoS$_2$ nanostructures such as QDs, nanosheets and nanorods have been observed at different T$_R$ ranging from 90 to 160 \degree C. On increasing synthesis temperature from 90 to 120 \degree C, increase in the number of QDs as well as nanosheets has been observed. Nanorod formation is observed on increasing T$_R$ from 120 to 160 \degree C. Further, optical emission and absorption spectra of these chemically synthesized nanostructures are found to be morphology dependent. Compositional analysis done using X-ray photoelectron spectroscopy (XPS) and Fourier transform infrared spectroscopy (FTIR) shows surface groups presence as well as MoS$_2$ formation in synthesized nanostructures.

\section{Experimental Section}
\begin{figure}[htbp]
\includegraphics[width=\linewidth]{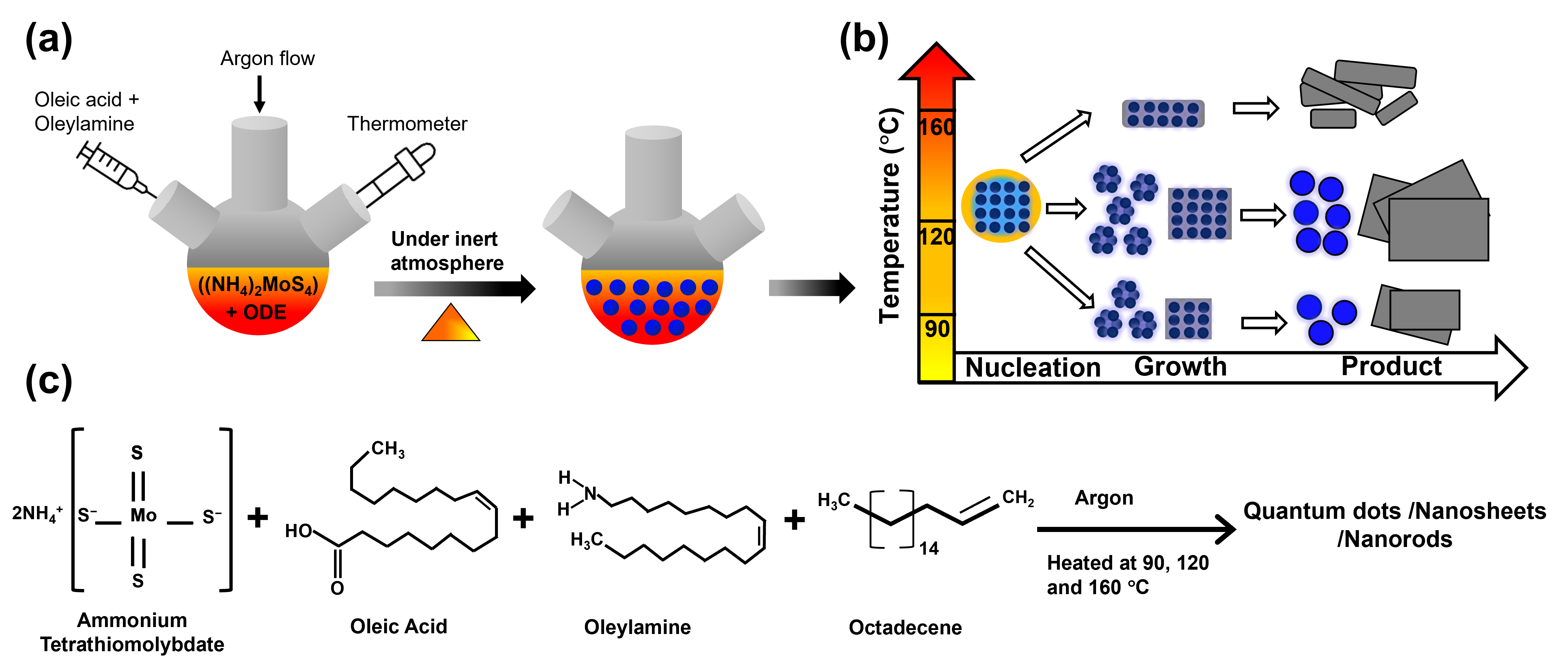}% Here is how to import EPS art
\caption{(a) Schematic illustration of colloidal synthesis of MoS$_2$ nanostructures using one pot heat-up method; (b) possible mechanism of nanostructures formation at different reaction temperatures; (c) chemical equation depicting reaction between (NH$_4$)$_2$MoS$_4$, oleic acid and oleylamine to form MoS$_2$ nanostructures.}
\label{Reaction_schematic}
\end{figure}
Figure \ref{Reaction_schematic}(a) shows the schematic illustration of MoS$_2$ nanostructures synthesis wherein a three-neck flask was used to mix all the reaction precursors. Ammonium tetrathiomolybdate ((NH$_4$)$_2$MoS$_4$) was used as a source of both molybdenum (Mo) and sulfur (S) elements with oleylamine (OAm) and oleic acid (OAc) as two capping ligands and 1-octadecene (ODE) as a high boiling point non-coordinating solvent. OAm and OAc act as stabilizing agents to quench the dangling bonds on the surface of nanostructures during growth process. The amount of 13 mg of (NH$_4$)$_2$MoS$_4$ was mixed with 1 ml of OAc, 3 ml of OAm, and 6 ml of ODE in the three-neck flask. Three mixtures were prepared in three separate flasks and were degassed in vacuum for 30 minutes with continuous stirring. All the mixed solutions were heated at 90, 120 and 160 \degree C respectively in argon atmosphere for 25 minutes. All heated solution were cooled down to room temperature and then centrifuged at 7000 rpm for 15 minutes. The obtained precipitates for all three temperatures were dispersed in chloroform solvent to avoid agglomeration of nanostructures. Due to high pyrolysis temperature of (NH$_4$)$_2$MoS$_4$ ($\approx$155 \degree C), we observe no nanostructure precipitation for T$_R$ $<$ 90 \degree C while a very fast uncontrollable reaction for T$_R$ $>$ 160 \degree C led to non-emissive black precipitate formation. The mechanism of resultant nanostructures prepared at 90, 120 and 160 \degree C is schematically explained in Figure \ref{Reaction_schematic}(b) showing formation of different shaped nanostructures (nanoparticles, nanosheets and nanorods) of MoS$_2$ with Figure \ref{Reaction_schematic}(c) showing the reaction equation with all precursors.

\section{Results and discussion}
\begin{figure}[htbp]
\centering
\includegraphics[width=10 cm]{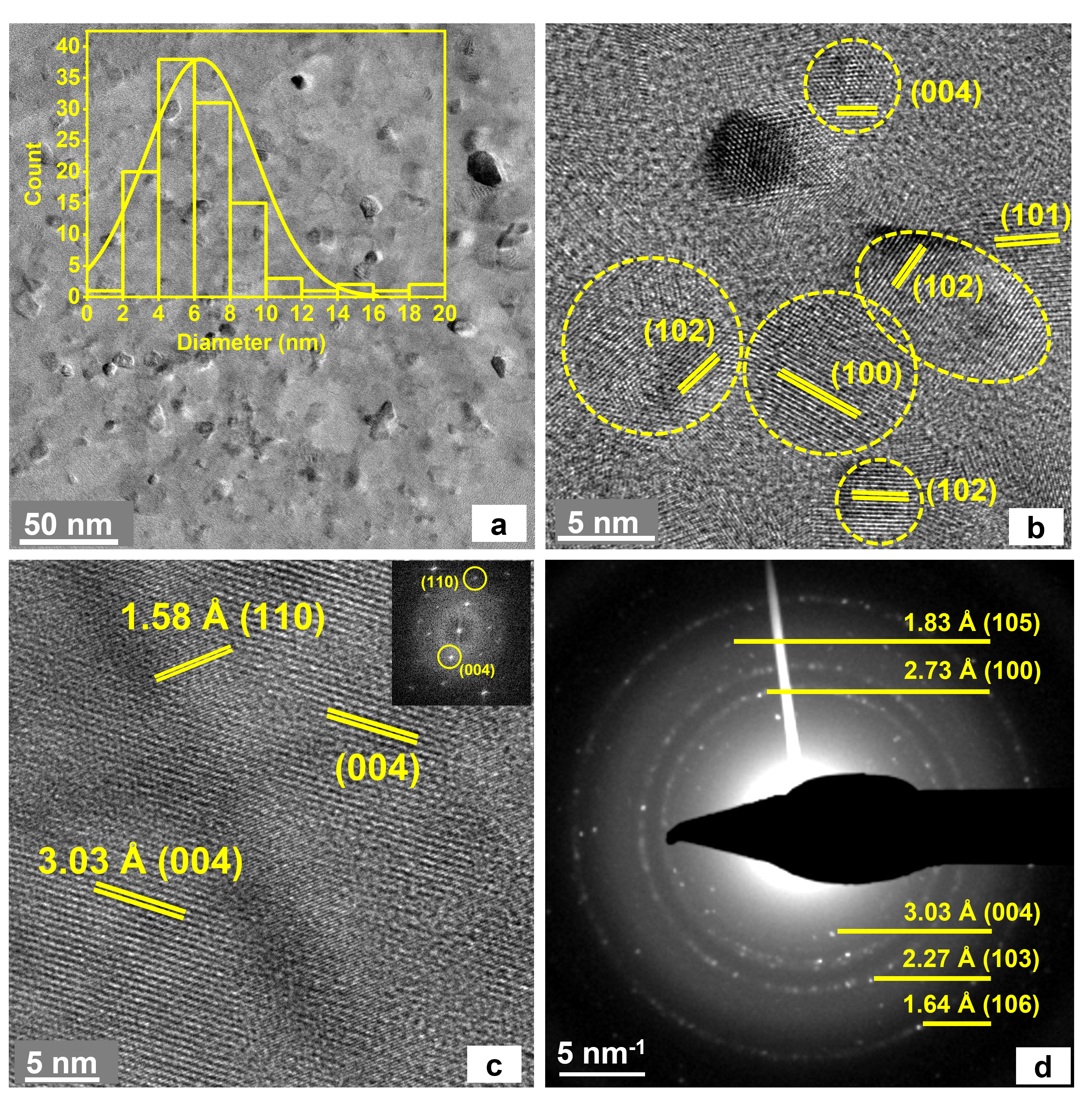}% Here is how to import EPS art
\caption{HRTEM images of nanostructures prepared at T$_R$ = 90 \degree C showing (a) quantum dots with hkl planes marked as well as size distribution histogram in inset showing average diameter, and (b) nanosheet with interatomic distances 3.03 \AA (004) and 1.58 \AA (110) with FFT in inset; (c) SAED pattern recorded for the QDs is showing orientation of QDs along different planes. The interatomic distances with corresponding (hkl) planes are marked for all points in SAED pattern.}
\label{90C}
\end{figure}

Structural analysis of as-synthesized MoS$_2$ nanostructures fabricated at different T$_R$ were investigated by transmission electron microscopy (TEM) and high resolution transmission electron microscopy (HRTEM) (Titan Themis 300kV from FEI). For a reaction temperature of 90 \degree C, MoS$_2$ QDs were predominantly formed with a gaussian size distribution and an average particle size of $\approx$5.6 nm in diameter evident from HRTEM image (Figure \ref{90C}a). The low T$_R$ (90 \degree C) led to small fraction of monomer formation ($<$ supersaturation state) for nucleation favouring QDs with a small probability of nanosheet formation. OA among QDs is favoured by capping agents (OAc and OAm) that get adsorbed on the nanoparticle surface by hindering the competitive Ostwald ripening (OR) process. QDs formed are not orientation specific as shown in Figure \ref{90C}a depicting (100), (102), (004) and other planes that may result in polycrystalline sheets formation through coalescence process\cite{thanh2014mechanisms} followed by OR evident in Figure \ref{90C}(b). Higher surface energy of nanocrystals is responsible for non-classical mechanism of aggregation forming attachment along the direction to reduce energy giving it orientation as is observed in growth along (004) and (110) plane in fast Fourier transform (FFT) of nanosheets in inset of Figure \ref{90C}(b). The polycrystalline nature of nanoparticles is confirmed in selected area electron diffraction (SAED) pattern shown in Figure \ref{90C}(c) by the presence of (004), (100), (103), (105), and (106) planes with the interplanar distance corresponding to 2-H MoS$_2$ that agrees with (hkl) planes calculated from HRTEM image.

\begin{figure}[t]
\centering
\includegraphics[width=10 cm]{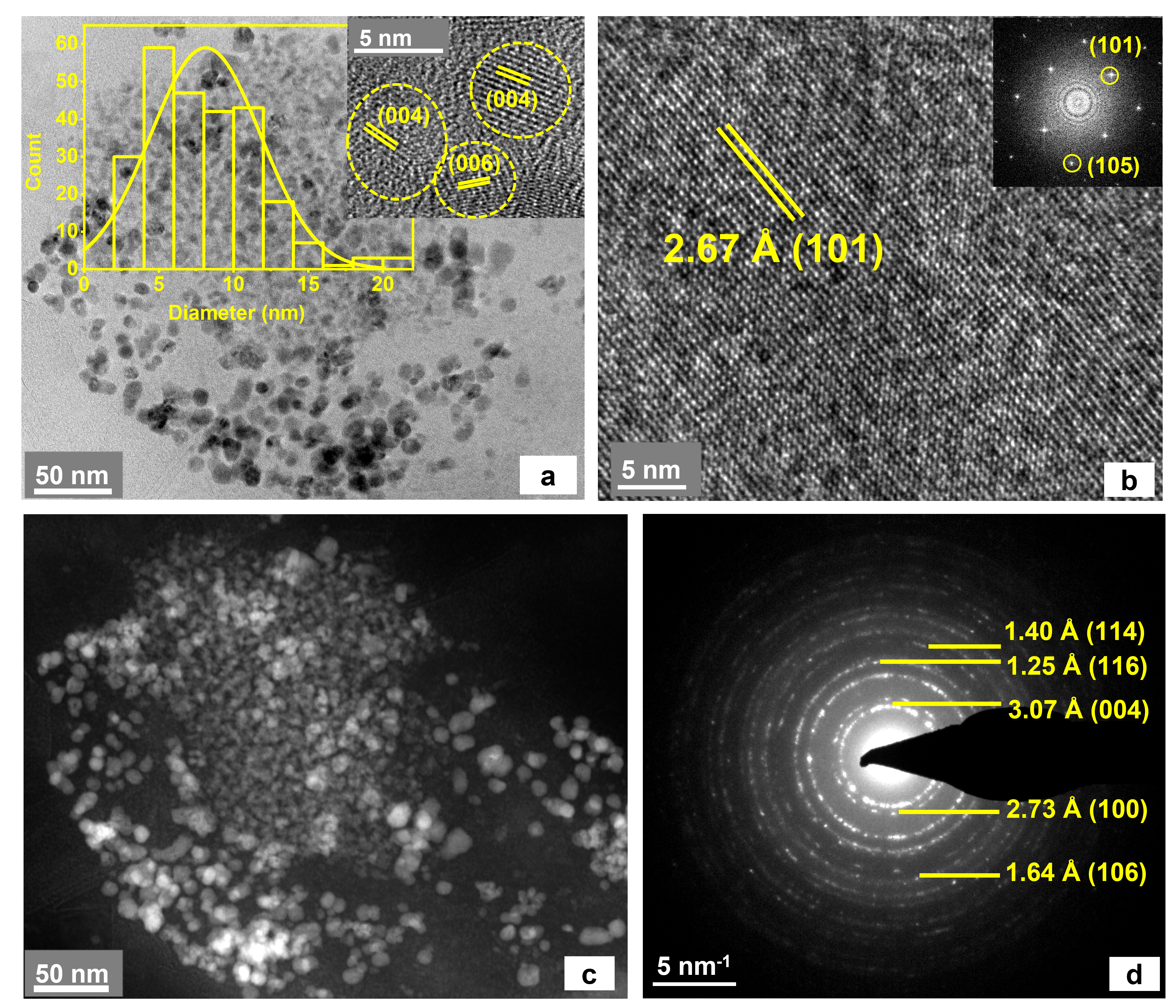}
\caption{(a) TEM image of MoS$_2$ QDs synthesized at T$_R$ = 120 \degree C is presented with hkl planes marked HRTEM image of QDs and size distribution histogram in insets; (b) HRTEM of nanosheet having an interatomic distance of 2.67 \AA (101) with FFT in inset; (c) dark-field STEM image of QDs; and (d) SAED pattern recorded for MoS$_2$ QDs showing multiple rings which is attributed to polycrystalline nature of synthesized QDs. All the rings are marked with respective interatomic distances and their corresponding (hkl) planes.}
\label{120C}
\end{figure}

\begin{figure}[b]
\centering
\includegraphics[width=10 cm]{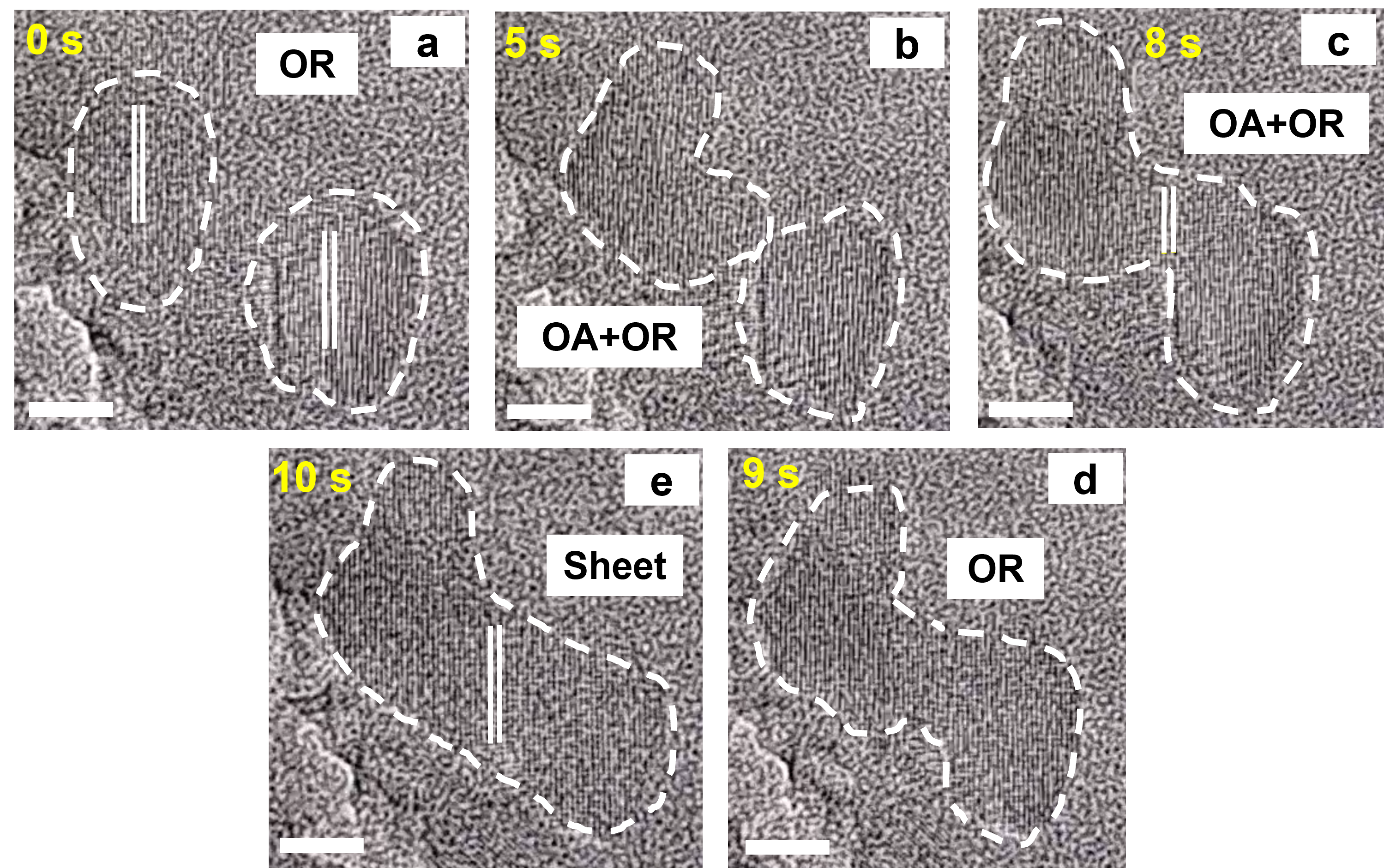}% Here is how to import EPS art
\caption{HRTEM images of the nanoparticles evolution with increasing reaction time showing the presence of (a) OR at 0s; (b) OA+OR at 5s; (c) growth after coalescing with OA+OR at 8s; (d) growth due to OR at 9s; and (e) final nanosheet formation at 10 s. Scale bar of HRTEM images taken at 300 KeV is 10 nm.}
\label{OA_Mech}
\end{figure}

MoS$_2$ sample prepared at 120 \degree C T$_R$ shows the presence of QDs as well as nanosheets (Figure \ref{120C} (a\&b)). At higher temperature (120 \degree C), the number of QDs and the nanosheet sizes formed is much larger as compared to the nanostructures formed at lower temperatures (90 \degree C). With increase in the synthesis temperature, the supersaturation (number of monomers) increases leading to more nucleation followed by growth and thus larger number of QDs precipitation take place. The size distribution histogram shows the average size of $\approx$8.1 nm in diameter of QDs (inset Figure \ref{120C}a). HRTEM image of polycrystalline sheet formed at 120 \degree C has been attached in Figure \ref{120C}(b). Dark-field Scanning transmission electron microscope (STEM) imaging was done for better contrast images as seen in Figure \ref{120C}(c). SAED pattern (Figure \ref{120C}d) shows polycrystalline nature of these QDs leading to formation of polycrystalline nanosheets. The interplanar distances calculated using SAED pattern are in good agreement with the interplanar spacings calculated from HRTEM images, indicating hexagonal lattice structure of MoS$_2$ nanostructures. A time-dependent TEM study of the nanosheet formation clearly shows a combination of OA and OR interplay to form a sheet as evident in Figure \ref{OA_Mech}(a-e). The growth of two nanoparticles with same orientation take place by Ostwald ripening process followed by coalescence\cite{thanh2014mechanisms} of these two nanoparticles through OA. After coalescing with OA and OR, the final structure forms and grows further by OR leading to formation of nanosheet as is shown in Figure \ref{OA_Mech}(e).

\begin{figure}[t]
\centering
\includegraphics[width=9 cm]{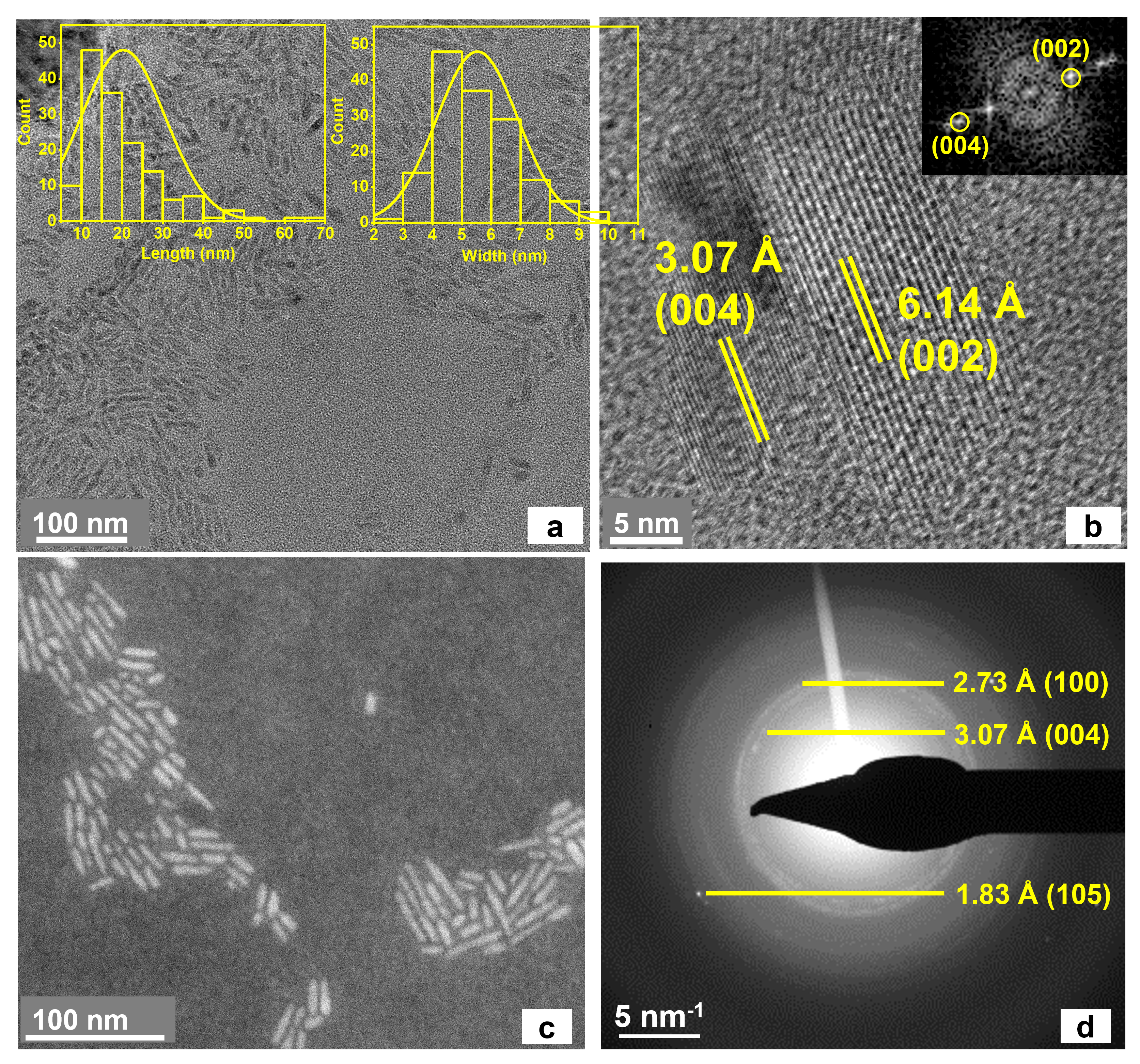}
\caption{(a) TEM image of nanorods with size distribution histogram in inset; (b) HRTEM image of nanorods growth along (002) and (004) planes with interplanar distances 3.07 \AA and 6.14 \AA respectively with FFT in inset; (c) dark-field STEM image of nanorods; and (d) SAED pattern of nanorods is showing presence of (100), (004), and (105) hkl planes. The interatomic distances are marked with their corresponding (hkl) planes.}
\label{160C}
\end{figure}

\begin{figure}[htbp]
\centering
\includegraphics[width=10cm]{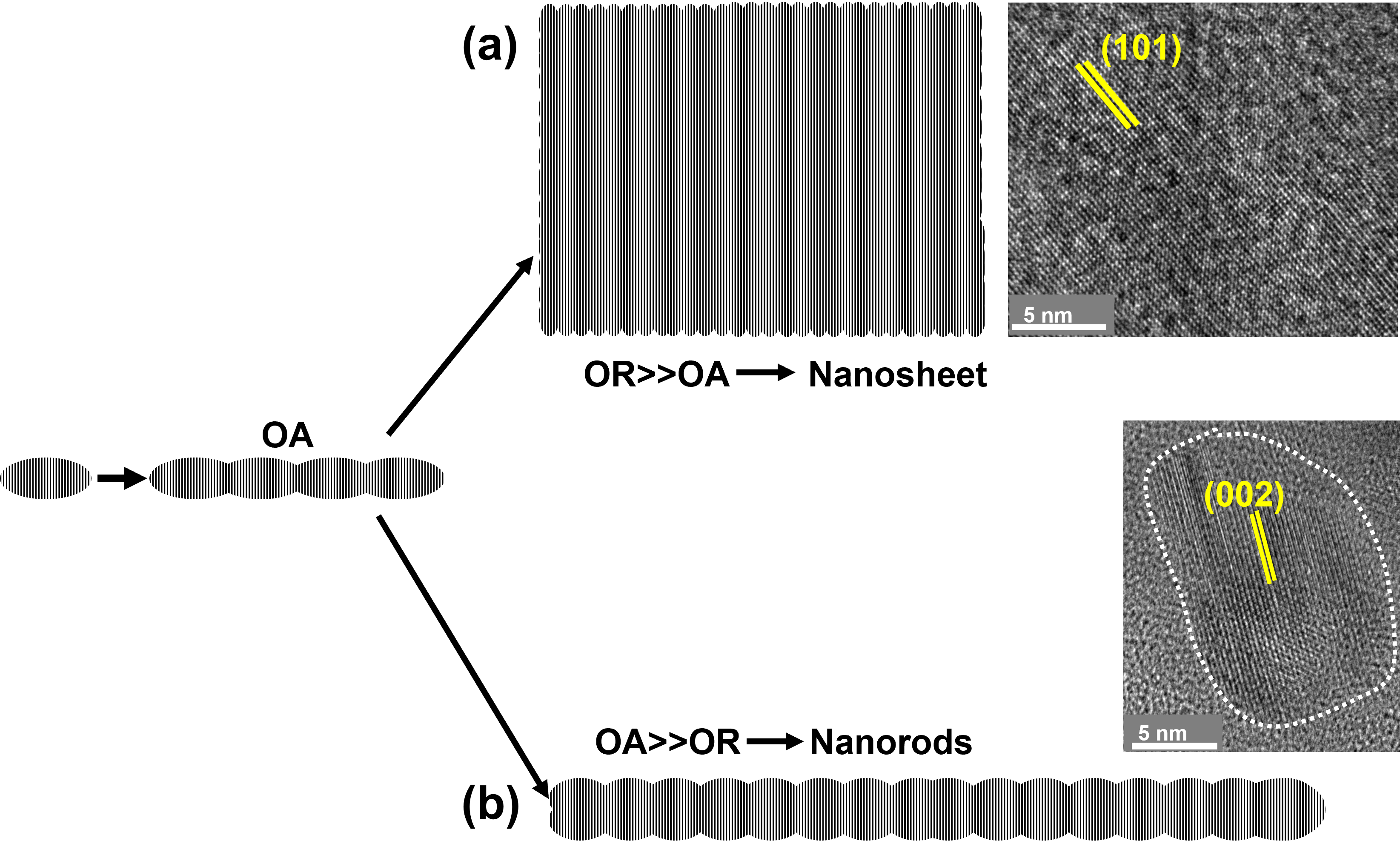}
\caption{Proposed schematic for different shaped nanostructures formation where nanoparticles attachment take place through OA process followed by (a) nanosheet formation when OR$>>$OA, and (b) nanorod formation when OA$>>$OR.}
\label{Proposed_mech}
\end{figure}

On further increasing the reaction temperature to 160 \degree C, nanorods formation was observed (Figure \ref{160C}). Energy supplied by high T$_R$ increases the monomer formation to the state of high supersaturation leading to generation of large number of nuclei. Based on the experimental results we hypothesize that OA is highly favoured over OR due to high temperature and fast reaction time leading to formation of nanorods as seen in HRTEM image of nanostructures in Figure \ref{160C}(b). The OA is corroborated by the growth along the plane (002) with an interatomic distance of $\approx$6.13 Å corresponding to MoS$_2$ (Figure \ref{160C}b). The average coverage area of these nanorods is $\approx $494 nm$^2$ with an average length of $\approx$42 nm which is shown in inset Figure \ref{160C}(a). STEM image showing clear nanorods with very good contarst is attached in Figure \ref{160C}(c). SAED pattern shows presence of three planes (004), (100), and (105) with interatomic spacings 3.07, 2.73 and 1.83 Å, respectively. The proposed mechanism of nanosheet and nanorods formation is shown in Figure \ref{Proposed_mech} wherein the dominance of OR leads to formation of nanosheets whereas a dominant OA leads to the formation of nanorods.

Compositional analysis was done to ascertain the stoichiometry of the formed nanostructures at  90, 120 and 160 \degree C temperatures. The energy dispersive X-ray spectroscopy (EDS) spectra as shown in supplementary Figure S1 with the dark-field STEM images on which EDS was recorded, confirming the formation of 1:2 (Mo:S) stoichiometric MoS$_2$ formation.

\begin{figure}[htbp]
\centering
\includegraphics[width=\linewidth]{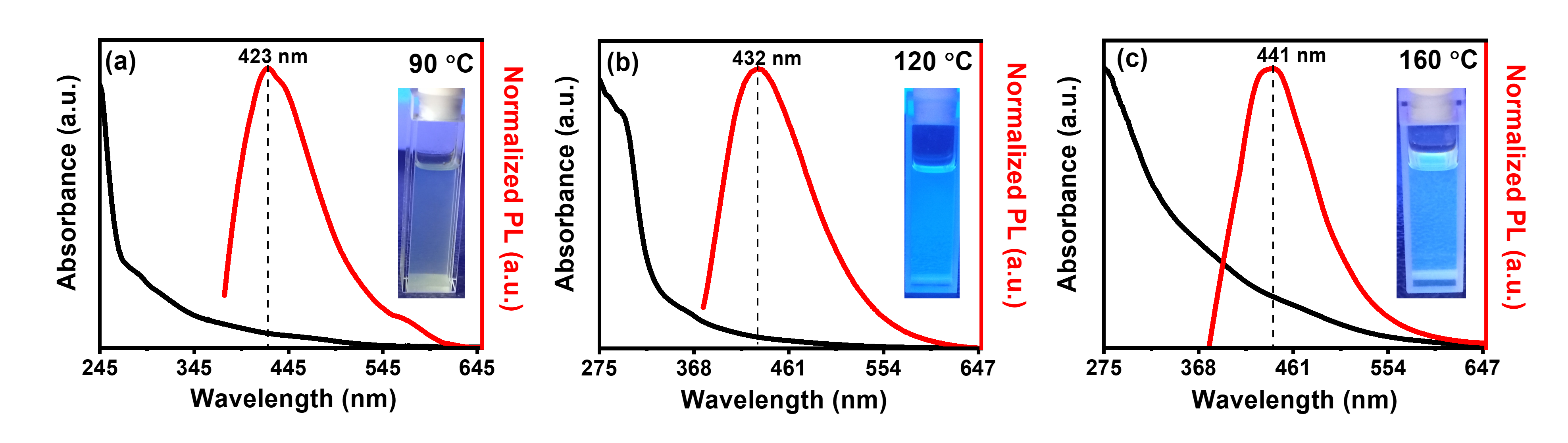}
\caption{Normalized absorbance and PL spectra with inset showing fluorescing colloidal MoS$_2$ nanostructures under 365 nm UV irradiation synthesized at (a) 90 \degree C, (b) 120 \degree C, and (c) 160 \degree C reaction temperatures.}
\label{abs}
\end{figure}

The optical properties of the prepared nanostructures were studied using UV-Vis spectroscopy and fluorescence spectroscopy. Absorption and PL spectra of nanostructures prepared at different temperatures are shown in Figure \ref{abs}. On increasing the T$_R$ from 90 to 160 \degree C, Urbach tail redshifts from 500 to 640 nm. The red shift in Urbach tail in absorption spectra can be explained by the change in the bandgap of the nanostructure with changing size and confinement with bandgap increasing from 4.7 eV in 90 \degree C to 3.9 eV in 120 \degree C to 3.67 eV in 160 \degree C. The direct bandgap was calculated by Tauc plot and is shown in Figure S2. The emission maxima of fluorescence spectra recorded at an excitation wavelength ($\lambda$$_{Ex}$) of 380 nm redshifts by $\approx$18 nm with increase in the T$_R$ from 90 to 160 \degree C as evident in Figure \ref{abs}. This redshift in emission maxima is attributed to increase in the size of nanostructures and changing morphology leading to change in bandgap. The images of MoS$_2$ nanostructures dispersed in chloroform were captured upon UV irradiation and are attached in inset of Figure \ref{abs} where 90 \degree C prepared nanostructures are poorly emitting while 120 \degree C and 160 \degree C prepared nanostructures are highly blue emitting.The absolute fluorescence QY was calculated using QuantaPhi Horiba Fluoromax plus spectrometer at different $\lambda$$_{Ex}$ and the values are shown in Table 1 for all the nanostructures synthesized at different T$_R$. From Table 1, we see zero QY for 90 \degree C prepared nanostructures while a maximum of 4.4\% QY value obtained for 120 \degree C prepared nanostructures at $\lambda$$_{Ex}$ = 440 nm and 3.06\% for 160 \degree C at $\lambda$$_{Ex}$ = 380 nm. The change in QY with $\lambda$$_{Ex}$ is attributed to the large size distribution. At lower temperature (90 \degree C), quantum dots and nanosheets are present in lesser number than at higher temperatures (120 \degree C) which was observed from TEM. The possible reason for zero QY value for 90 \degree C prepared nanostructures can be more non-radiative recombination of excitons due to presence of defects or surface states\cite{lambora2023role}. With increase in reaction temperature to 160 \degree C, the QY is smaller than QDs (120 \degree C) which is due to change in morphology from QDs to nanorods where charge carriers are poorly confined in comparison to QDs. The variation of QY with $\lambda$$_{Ex}$ is shown in Figure S3. Decay lifetime was also measured using DeltaFlex TCSPC Lifetime Fluorometer for nanostructures prepared at all three T$_R$ and the lifetimes curves are shown in Figure S3. Average decay lifetimes of nanorods and QDs synthesized at 160 and 120 \degree C are 2.03 ns and 2.32 ns respectively, have smaller average lifetime than 90 \degree C prepared QDs (lifetime = 2.83 ns) that is attributed to the defect/surface states presence\cite{xu2019fabrication}. 

\begin{table}[htbp]
\centering
\caption{\bf QY (\%) variation with $\lambda$$_{Ex}$ and decay lifetimes at different T$_R$}
\begin{tabular}{|c|c|c|c|c|c|c|c|c|}
\hline
 & \multicolumn{6}{c}{\bf $\lambda$$_{Ex}$ (nm)}& & \\
\hline
\bf T$_R$ (\degree C) & \bf 360 & \bf 380 & \bf 400 & \bf 420 &  \bf 440 &  \bf 460 & \bf 480 & \bf Lifetime(nm) \\
\hline
$90$ & $0$ & $0$ & $0$ & $0$ & $0$ & $0$ & $0$ & $2.83$ \\
\hline
$120$ & $3.37$ & $3.99$ & $3.48$ & $3.75$ & $4.4$ & $3.96$ & $3.35$ & $2.32$ \\
\hline
$160$ & $2.26$ & $3.06$ & $2.87$ & $2.9$ & $2.63$ & $1.88$ & $1.56$ & $2.03$ \\
\hline
\end{tabular}
\label{tab:Table 1}
\end{table}

\begin{figure}[t]
\centering
\includegraphics[width=10cm]{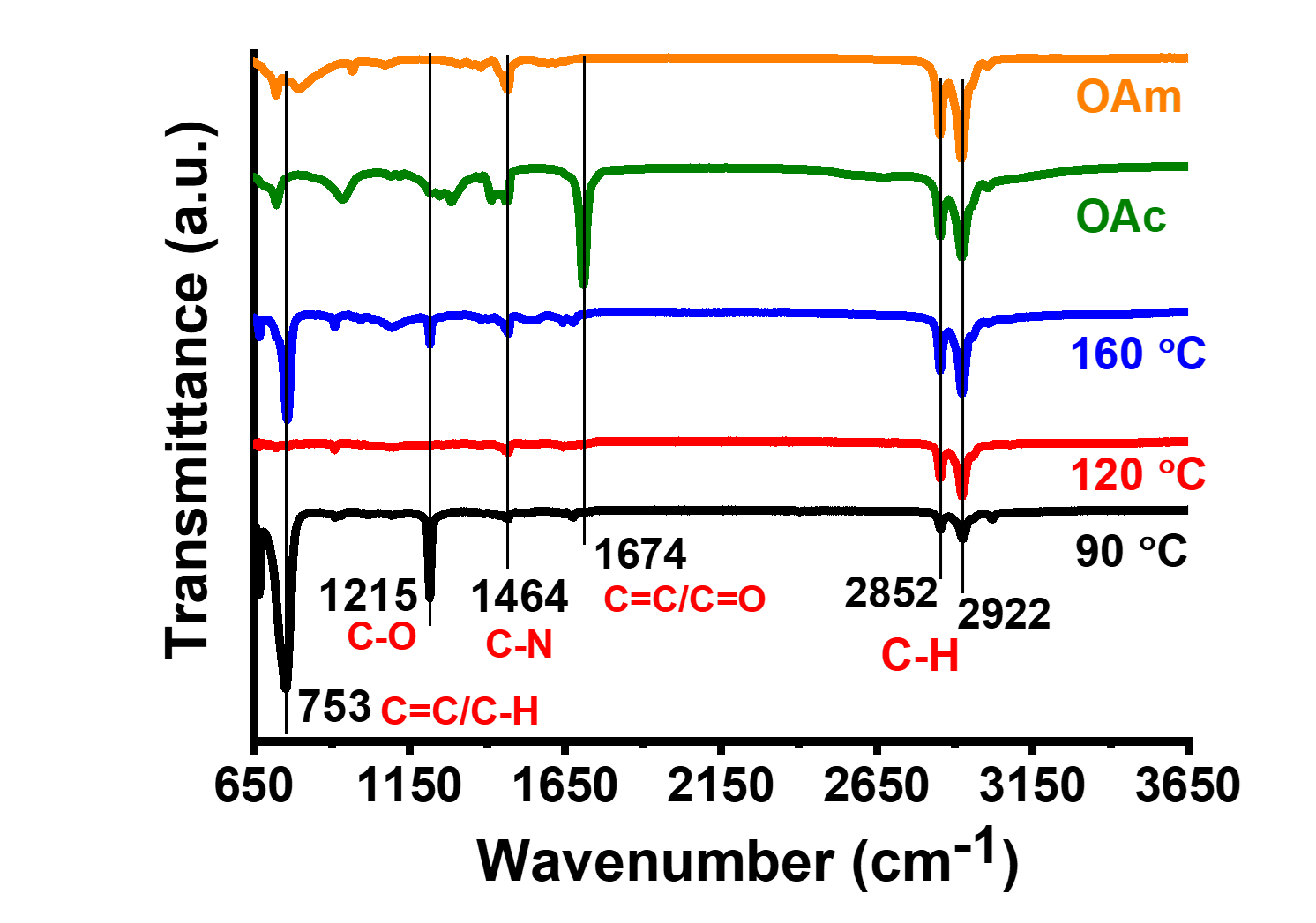}% Here is how to import EPS art
\caption{Comparison in FTIR spectra of the synthesized nanostructures at varying temperature of 90 \degree C, 120 \degree C, 160 \degree C and ligands OAm \& OAc. The common peaks show the ligands presence on MoS$_2$ nanostructures surface.}
\label{ftir}
\end{figure}

\begin{figure}[b]
\centering
\includegraphics[width=\linewidth]{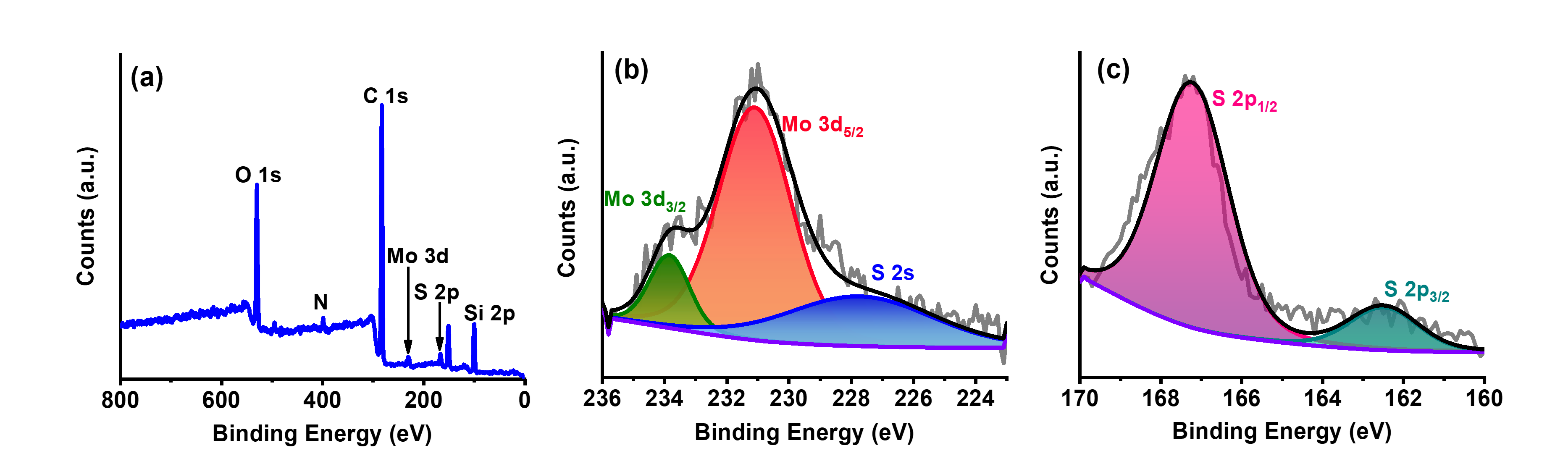}% Here is how to import EPS art
\caption{(a) Wide scan XPS spectrum of MoS$_2$ nanostructures synthesized at 120 \degree C; (b) deconvoluted Mo 3d XPS spectrum showing Mo 3d$_{5/2}$, Mo 3d$_{3/2}$ and S2s; (c) deconvoluted S 2p XPS spectrum presenting S 2p$_{1/2}$ and S 2p$_{3/2}$ phases.}
\label{XPS}
\end{figure}

Further information about the surface groups was harnessed from Fourier transform infrared spectroscopy (FTIR) (Figure \ref{ftir}). FTIR spectra of pure OAm and OAc was also recorded to confirm the surface groups presence. In Figure \ref{ftir}, the surface groups C=C, C=O, C-N, and C-H confirm the OAc and OAm presence on the MoS$_2$ nanostructure surface synthesized at 90, 120, and 160 \degree C reaction temperatures. The zoomed-in FTIR spectra in ranges of 4000-1200 cm$^{-1}$, 1200-600 cm$^{-1}$, and 600-420 cm$^{-1}$ are shown in Figure S4 with all the bonds marked at their respective wavenumbers. In Figure S4(c), the peak at 537 cm$^{-1}$ correspond to Mo-S stretching vibration. The FTIR results confirm presence of surface groups in the synthesized nanostructures.

X-ray photoelectron spectroscopy (XPS) measurements were done to confirm the elemental composition of MoS$_2$ nanostructures synthesized at 120 \degree C and are shown in Figure \ref{XPS}. Wide scan XPS spectrum (Figure \ref{XPS}a) shows the presence of molybdenum (Mo) and sulphur (S) along with carbon (C), nitrogen (N) and oxygen (O). C, N and O elemental presence is due to the capping agents (OAm and OAc). Silicon (Si) peak is also observed which is attributed to the Si substrate used for the measurement over which the colloidal nanostructure solution was drop-casted. The deconvoluted Mo 3d spectrum (Figure \ref{XPS}b) shows the presence of Mo 3d$_{3/2}$ and Mo 3d$_{5/2}$ positioned at 233.9 eV and 231.1 eV respectively. These peaks are blue shifted in comparison to bulk MoS$_2$ (3d$_{3/2}$ $\approx$230 eV and 3d$_{5/2}$ $\approx$233 eV) by the order of $\approx$2 eV. Similar observation was found for deconvoluted S 2p spectrum as well. In Figure \ref{XPS}c, S 2p$_{1/2}$ and S 2p$_{3/2}$ peaks are positioned at 167.3 eV and 162.5 eV which are also blue shifted in comparison to bulk MoS$_2$ in which S 2p$_{1/2}$ and S 2p$_{3/2}$ peaks appear at $\approx$162.9 eV and $\approx$164.1 eV. The increase in the binding energy values for both Mo and S can be attributed to quantum confinement effects in nanostructures\cite{lambora2023role}.

\section{Conclusion}
We report the first time study of shape transformation of MoS$_2$ colloidal nanostructures from QDs to nanosheets to nanorods with variation in the synthesis reaction temperature from 90 to 160 \degree C. OA of highly energetic facets of quantum dots leads to nanosheet and nanorod formation via Ostwald ripening. The nanosheet formation has been explained through time-dependent TEM images showing coalescence through oriented attachment followed by Ostwald ripening of quantum dots. The interplanar distances calculation using SAED pattern are in good match with the bulk MoS$_2$ confirming no lattice structure transformation for the synthesized nanostructures. The compositional study using EDS and XPS shows the Mo and S elemental presence in the prepared nanostructures. FTIR results have confirmed the presence of surface groups such as C=C, C=O, C-N, O-H, N-H, C-H due to capping ligands and Mo-S due to MoS$_2$ in the synthesized nanostructures. From the optical characterization, we have shown the high emission from quantum dots and nanosheets prepared at 120 \degree C with quantum yield of 4.4\% than 90 \degree C (quantum dots + nanosheets) and 160 \degree C (predominantly nanorods). Further study is ongoing to understand the effect of synthesis reaction time on morphology and the optical properties of MoS$_2$ nanostructures for temperatures 120 and 160 \degree C.

\section*{Acknowledgments}
The authors would like to thank MNCF facility present at CeNSE, IISc Bangalore, India. We acknowledge ISRO (grant number- ISTC/PIA/AB/460) and Ministry of Electronics and Information Technology (MeitY) (SP/MITO-19-0005.05) for the financial support.

%Bibliography
\bibliographystyle{unsrt}  
\bibliography{references}

\end{document}